\documentclass[preprint]{aastex62}

\usepackage{amsmath,amstext}

\graphicspath{{./}{figures/}}

\submitjournal{ApJ}

\shorttitle{Metallicity and Convective Overshoot}
\shortauthors{Dornan \& Lovekin}

\begin{document}

\title{The Effects of Metallicity on Convective Overshoot Behaviour in Models of $\delta$ Scuti Variable Stars}

\author{V. Dornan}
\affil{Department of Physics and Astronomy, McMaster University, Hamilton, ON,  Canada}
\affiliation{Physics Department, Mount Allison University, Sackville, NB, Canada}

\author{C.C. Lovekin}
\affiliation{Physics Department, Mount Allison University, Sackville, NB, Canada}
\email{clovekin@mta.ca}

\begin{abstract}

$\delta$ Scuti variables are stars which exhibit periodic changes in their luminosity through radial and non-radial pulsations. Internally, these stars have relatively small convective cores, and convective overshoot can significantly affect the size.  Recently, models of radial pulsation in $\delta$ Scuti stars found a strong correlation between the pulsation constant ($Q$) as a function of effective temperature and the amount of convective overshoot within the star. However, only models with metallicities of $Z=0.02$ were examined, leaving the dependence of this relationship on chemical composition unknown. In this work, we have extended the model grid to cover a range of metallicities using \texttt{MESA}, and analyzed the models’ pulsation properties using \texttt{GYRE}. By varying the models’ mass, rotation speed, convective overshoot, and metallicity, we studied the behaviour of $Q$ at low temperature. We found that the updated convective boundary treatment in MESA changes the overshoot dependence found previously, and the value of the slope depends on both rotation and overshoot.  We also found that there is a metallicity dependence in the $Q$ values. The lowest metallicity models in our grid reached higher temperatures than previously studied, revealing a parabolic relation between $\log Q$ and $\log T_{eff}$. 

\end{abstract}


\section{Introduction}

\label{sec:intro}

$\delta$ Scuti variables are located on the pre-main sequence to the post-main sequence with masses between 1.4 and 2.5 M$_{\odot}$, located where the classical instability strip intersects the main sequence. They pulsate in either low order pressure modes ($p$ modes) \citep{xiong2016} or mixed modes with both $p$ and $g$ mode characteristics \citep{guzik2000,Dupret2004,dupret2005}.  The periods of these modes are shorter than those of classical Cepheids, with periods ranging from tens of minutes to $\sim$ 5 hours. These periods are short enough to be easily studied using ground based data, however space-based missions such as Kepler \citep{basri} and TESS \citep{ricker} have produced extremely rich pulsation spectra with large numbers of identifiable frequencies.  These observations provide enough frequencies to make echelle diagrams \textbf{possible} for some of these stars particularly those closer to the ZAMS. When echelle diagrams are possible, the large separation, and in some cases the rotation rate can be determined for these stars \citep{paparo2016,Bedding2020,Murphy2020}.  The large separation, combined with fitting of individual frequencies, can subsequently be used to constrain the convective core overshoot of these stars \citep{khalack2019}.  

This class of pulsating stars provides an intriguing theoretical challenge.  Theoretically, all stars in the $\delta$ Scuti instability strip are expected to undergo pulsations, while observations show that less than half of them do \citep{balona2011,balona2015}. Stable stars in the instability strip may be the result of nonlinear mode coupling, which can act to stabilize the pulsations \citep{dziembowski88,dziembowski1992}, or it has been proposed that the opacity driving mechanism is saturated  \citep{nowakowski05}.  Testing either of these proposals would require nonlinear models of non-adiabatic non-radial pulsations, which do not currently exist.

Some $\delta$ Scuti stars also show pulsations typical of the slightly lower mass $\gamma$ Doradus stars \citep{hareter2010,grigahcene2010,bradley2015}. $\gamma$ Doradus stars typically pulsate in $g-$modes, and have periods between 0.3 and 3 d, making them difficult to study with ground-based observations. Space-based observations have shown that there are far more hybrid stars than predicted by theory \citep{uytterhoeven2011}, and some authors have even suggested that all $\delta$ Scuti stars are hybrid stars \citep{balona2014, balona2015}.   

The distribution of hybrid stars is related to the presence of convection. The red edge of the $\delta$ Scuti instability strip is strongly dependent on the time-dependent treatment of convection. $\delta$ Scuti pulsations are mostly driven by instabilities in the HeII ionization zone \citep{chevalier1971}, and at the red edge the surface convection extends below this layer, damping pulsation. At the same time, the $\gamma$ Doradus stars are driven by a convective blocking mechanism \citep{guzik2000,Dupret2004}, and so convection is crucially important in determining where these stars will be unstable.

Studies of convection in $\delta$ Scuti stars are further complicated by rotation.  $\delta$ Scuti stars are often rapid rotators, with rotation rates as high as 200 to 250 km s$^{-1}$ \citep{breger2007} and a mean rotation velocity of $\sim$ 120 km s$^{-1}$ \citep{glebocki2005}.  Rotation affects the small-scale dynamics of convection, particularly in shear layers at the convective boundaries.  On larger scales, rotation can produce extra mixing that effectively increases the size of the convective core, extending the main sequence lifetime of a star. Both large and small scale effects are parameterized in stellar models using convective core overshoot. Although it is clear that some convective overshoot is required in stellar models to match observations, it is not clear how much is required, and whether it is the same in all stars \citep[see for example][and references therein]{Claret2018, Pedersen2018,Johnston2021}.  The details of both rotation and convective overshoot are key components to correctly predicting mode excitation in $\delta$ Scuti stars.

The $\gamma$ Doradus and $\delta$ Scuti stars are slightly more massive than the Sun, and so have both convective cores and shallow convective envelopes. The presence of both these convective zones and a large number of frequencies make them excellent laboratories for studying convective overshoot \citep{grigahcene2010}. 

Many types of variable stars' pulsations, including $\delta$ Scutis, are known to adhere to a period-mean density relation \cite[see, for example,][]{Cox-80}. Through this relation the period of a variable star's pulsations and the star's mean density can be related to yield a pulsation constant which can help compare stars of similar size and structure:
\begin{equation} \label{p-md}
    Q = \Pi \Big(\frac{\overline{\rho}}{\overline{\rho_{\odot}}}\Big)^{\frac{1}{2}} \\
\end{equation}
In this equation $Q$ represents the pulsation constant, $\Pi$ is the period of the star's pulsation in days, $\overline{\rho}$ is the mean density of the variable star, and $\overline{\rho_{\odot}}$ is the mean density of the Sun. The pulsation constant can be used to identify modes based on relationships determined by \citep{breger1990}. These relationships are calibrated to the Sun, and thus neglect the effects of both rotation and convective overshoot.  Without these effects, the applicability to $\delta$ Scuti stars is limited.

The relationship between the pulsation constant of $\delta$ Scuti stars as a function of effective temperature and their convective overshoot was investigated previously by  \cite{Fitch-81}. In this work Fitch was able to identify several interpolation formulae which could be used to predict the pulsation constants of $\delta$ Scuti stars. The formula provided in \citet{Fitch-81} are linear for solar metallicity models with $3.92 \geq \log(T_{eff}) \geq 3.83$. However, the model grid used to generate these interpolation formulae did not include any models with non-solar metallicity, no models with $\log(T_{eff}) \geq 3.92$, and only two models with $\log(T_{eff}) \leq 3.84$.

More recently, \cite{Lovekin-17} studied the behaviour of convective overshoot in models of $\delta$ Scuti and $\gamma$ Doradus stars between 1.2 M$_{\odot}$ and 2.2 M$_{\odot}$.
They found a strong correlation between the models pulsation constants as a function of effective temperature and their convective overshoot. The relationship between $\log Q$ and $\log T_{eff}$ could be described using a piece-wise linear function, with a break point at $\log(T_{eff}) \approx 3.83$. For temperatures higher than this point the pulsation constant is truly constant, as was found by \cite{Fitch-81}, while for temperatures lower than this point the slope of the function, which will be referred to as $m_{low}$, varies with the convective overshoot of the model. The relationship between the behaviour of $m_{low}$ and the 
exponential convective overshoot parameter ($f_{ov}$) as defined in \citep{Herwig2000} can be described with a linear fit:
\begin{equation} \label{eq:orig}
    m_{low} = (12 \pm 2) f_{ov} - (2.8 \pm 0.1) \\
\end{equation}

\cite{Lovekin-17} investigated the cause of this variation in the pulsation constants, and found a correlation with the normalized depth of the convective envelope. For models hotter than $\log T_{eff} = 3.83$, the surface convection zone depth remained roughly constant, accounting for a few percent of the stellar radius.  Below this temperature, the surface convection zone became progressively deeper as the temperature decreases, accounting for 30 to 40 \% of the stellar radius in the coolest models. The temperature at which the convection zones begin to expand, $\log T_{eff} = 3.83$, corresponds to the break point found in the $\log Q-\log T_{eff}$ relationship. \cite{Lovekin-17} proposed that $m_{low}$ could be measured for radially-pulsating $\delta$ Scuti stars, and could be used to constrain the average convective overshoot parameter for a sample of stars.

However, \cite{Lovekin-17} used only models of solar-like composition , that is to say $Z=0.02$ with the \citep{GS98} abundances. The effect of metallicity on the behaviour of $\log Q$ remains unclear.  Understanding the metallicity dependence of this relation will increase the applicability of this method to observed $\delta$ Scuti stars, and has the potential to vastly increase the number of stars with known convective core overshoot parameters.  A better understanding of convective core overshoot will in turn improve our understanding of a variety of problems in stellar and galactic astrophysics.

In this work, we extend the work of \cite{Lovekin-17} to a range of metallicities in order to determine the metallicity dependence of their $\log Q - \log T_{eff}$ relations.  In Section \ref{sec:models} we describe the details of the model grid. In Section \ref{sec:results} we describe our findings based on a full grid covering a range of metallicities, and discuss the effects of convection in Section \ref{sec:convection}. Finally we summarize our conclusions in Section \ref{sec:conclusion}.

\section{Models}
\label{sec:models}

We calculated a grid of stellar models using \texttt{MESA} (Modules for Experiments in Stellar Astrophysics) version 12115 \cite{Paxton-10,Paxton-13,Paxton-15,Paxton-17,Paxton-19}.  These models included the effects of convective overshoot in models from 1.2 to 2.2 M$_{\odot}$ in 0.1 M$_{\odot}$ increments.  Rotation rates were defined on the Zero Age Main Sequence (ZAMS) as a fraction of the critical velocity, ranging from 0.0 to 0.4 in steps of 0.1, replicating the original model grid of \cite{Lovekin-17}.  In this work, we extended the metallicity of the original model grid, and calculated models with metallicities of Z = 0.04, 0.03, 0.02, 0.01, 0.005, and 0.001.  All models were scaled-solar composition based on the \cite{GS98} metallicity. Other mixtures \citep{[eg.][]Asplund2005} require lower Z for the solar metallicity, and have lower abundances of some elements, including C and O.\texttt{MESA} uses the OPAL equation of state \citep{OPAL} supplemented with the SCVH tables \citep{SCVH} at low temperatures and densities.  Our full set of parameters at each metallicity is summarized in Table \ref{tab:1}.  

Models were calculated using the standard mixing length theory (MLT) of convection \citep{Bohm-58}, with a mixing length of $\alpha = 2.0$.  Convective overshoot was included at all convective boundaries using the exponential overshoot model \citep{Herwig2000}
\begin{equation}\label{eq:diffuse}
    D_{ov} = D_0 \exp\left(\frac{-2r}{f_{ov}H_P}\right).
\end{equation}
Here, $D_0$ is the diffusion coefficient, calculated by \texttt{MESA} using MLT at the convective boundary, $r$ is the radial distance within the star from the core boundary, $f_{ov}$ is the convective overshoot parameter, and $H_P$ is the pressure scale height at the core boundary. The convective overshoot parameter in Equation \ref{eq:diffuse} is allowed to vary between 0 and 0.1 for these models in steps of 0.02, again replicating the model grid of \citet{Lovekin-17}. This parameter determines the amount of convective overshooting present in the stellar model both above convective cores and below convective envelopes. Although these can be set separately, the same value was used for all convective regions in this work.

\texttt{MESA} also allows for variation in the rotation rate of the models. Rotation is included by imposing a surface rotation rate and forcing the ZAMS model to be uniformly rotating. \texttt{MESA} uses a shellular approximation for rotation \citep{meynet1997}, which allows it to include rotation in a pseudo-1D fashion. The shellular approximation breaks down at high rotation rates, but should be sufficient for the rotation rates considered here \citep{meynet1997}.  The rotation rate of the models is parameterized as $\Omega/\Omega_c$, where $\Omega$ is the surface angular velocity and $\Omega_c$ is the critical rotation. Rotation rates given in Table \ref{tab:1} are specified on the ZAMS, and rotation is allowed to evolve with time.  During evolution, transport of angular momentum is implemented via diffusion \citep{Paxton-13}.  

Models were evolved from the ZAMS to the end of the main sequence.  Detailed models were saved every 30 times steps.  The time steps were variable, and were determined by requiring the relative change in the structure (density, temperature, pressure) to be less than 10$^{-4}$ from one time step to the next.  

\begin{deluxetable}{ccc}
\caption{\label{tab:1} Summary of the parameters in our model grid. }
\tablehead{\colhead{Parameter} & \colhead{Range} & \colhead{Step} }
          \startdata
         Masses ($M_\odot$) & 1.2-2.2 & 0.1 \\
         Convective Overshoot ($f_{ov}$) & 0.0-0.1 & 0.02 \\
         Rotation Rates ($\Omega/\Omega_c$) & 0.0-0.4 & 0.1 \\
        \enddata
\end{deluxetable}


For each of the detailed models, we calculated pulsation frequencies using \texttt{GYRE} \citep{Townsend-13,Townsend2018}.  \texttt{GYRE} is able to solve both adiabatic and non-adiabatic pulsation equations for stellar models using a Magnus Multiple Shooting (MMS) numerical scheme. We performed adiabatic calculations to determine the $\ell = 0$ modes between $2.0 - 30.0$ d$^{-1}$.  This covers the first 6-7 radial orders for each model, however our analysis will focus on the radial fundamental mode.  \texttt{GYRE} also includes the effects of rotation on pulsation frequencies using the Traditional Approximation \citep{Eckart1960, lee1987}. This approximation is applicable to g-modes, but not the p-modes studied here. For this reason, we do not include the effects of rotation in \texttt{GYRE}, taking into account only the centrifugal distortion from the underlying \texttt{MESA} models.

\section{Behaviour of Q}
\label{sec:results}

\begin{figure}[h!tb]
    \centering
    \plotone{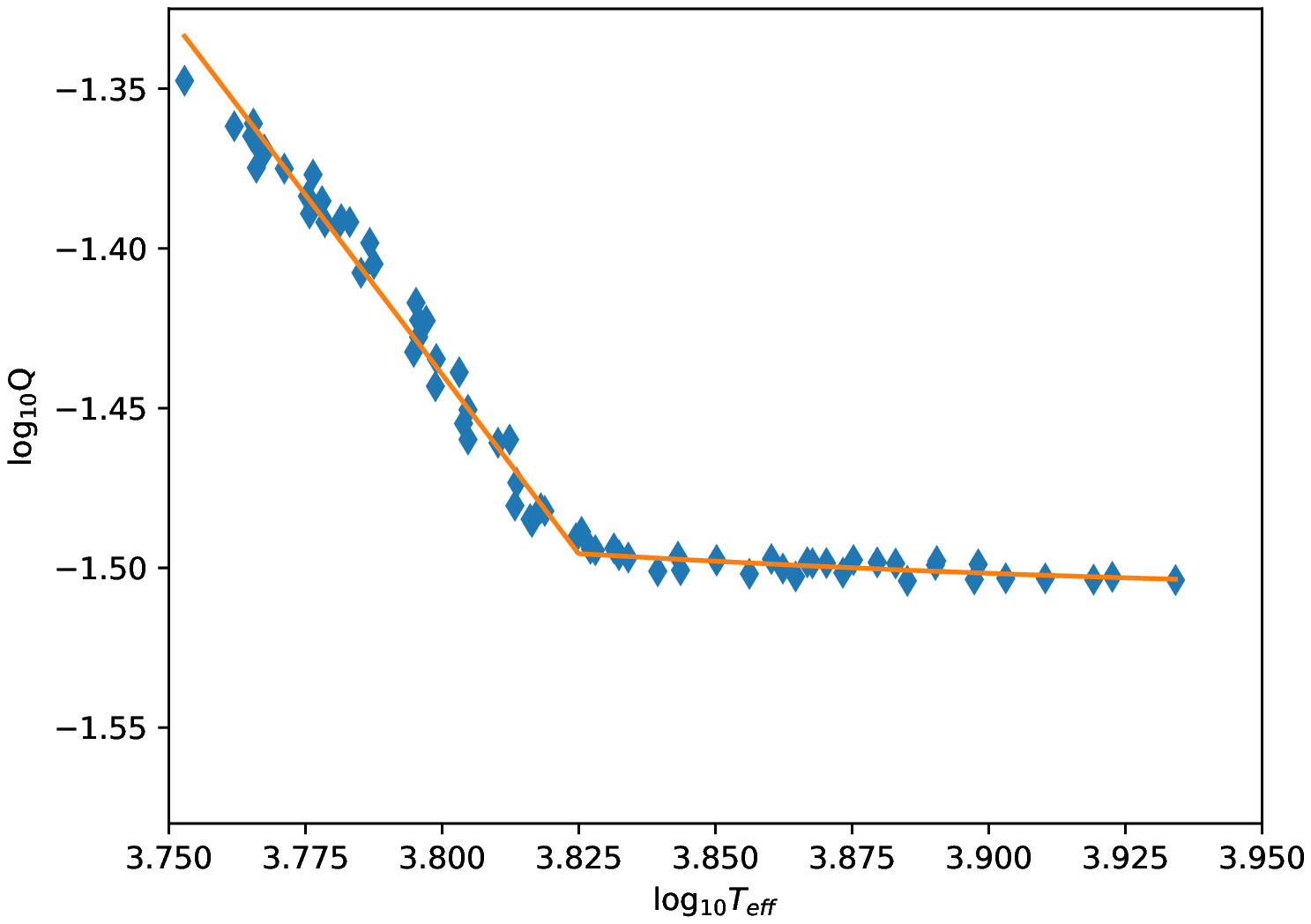}
    \plotone{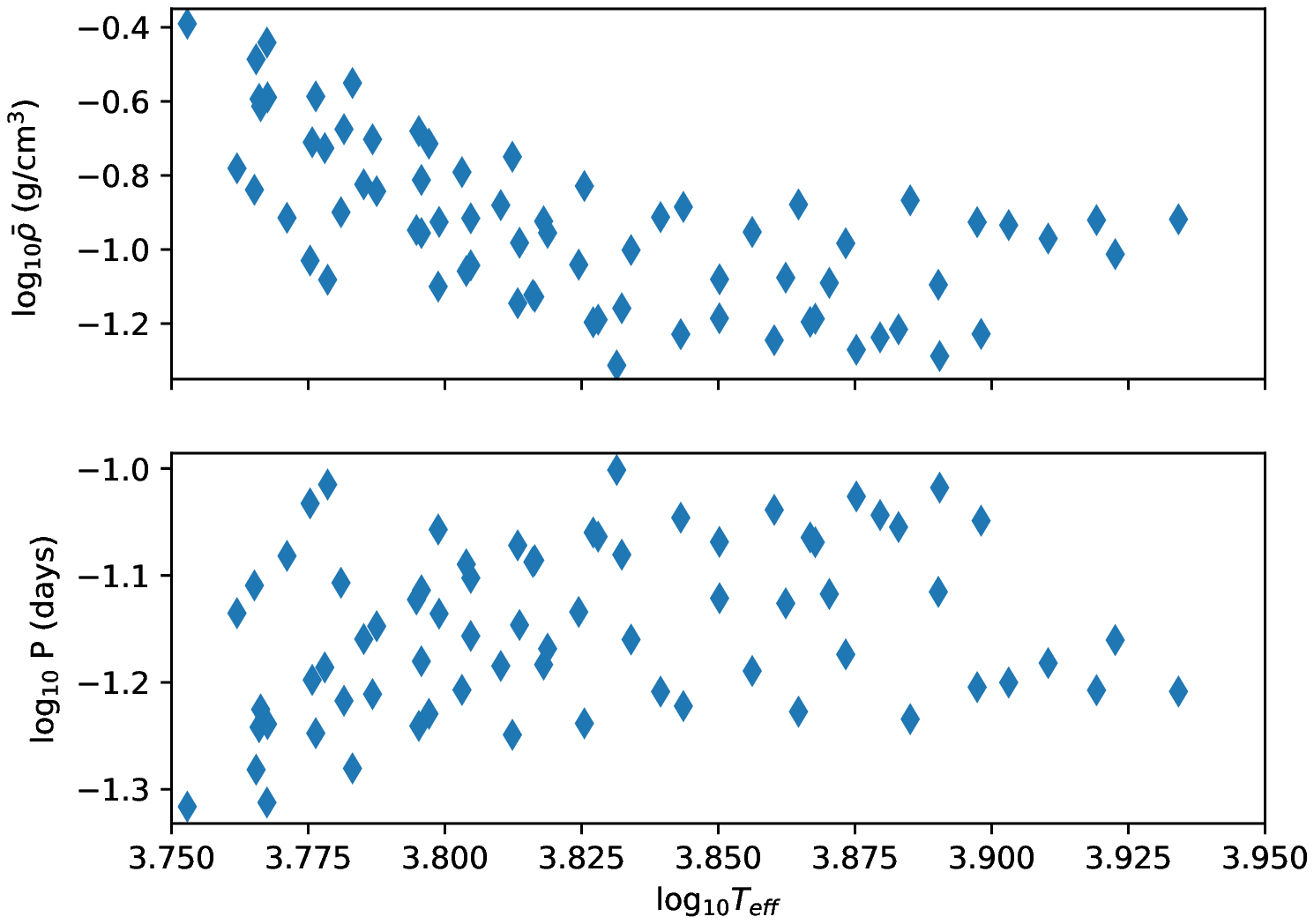}
    \caption{Top: Calculated $\log(Q)$ values (blue diamonds) for models at $Z = 0.04$ with $f_{ov} = 0$ and $\Omega/\Omega_c = 0$ plotted against $\log(T_{eff})$. The solid line shows the best fit to the data using a piece wise function.  The coefficients of both the linear and parabolic sections were free parameters, as given by Equation \ref{eqn:fitfunc}. Bottom: The behaviour of the log of the mean density and Period as a function of temperature for the same model.  }
    \label{piecewise}
\end{figure}

We calculated $Q$ values for the radial fundamental mode for each main sequence model in our grid using Equation \ref{p-md}.  We then plotted the $\log Q$ for each model versus $\log T_{eff}$, as shown in Figure \ref{piecewise} for the Z = 0.04 models with no rotation or overshoot.  As was found previously by \cite{Lovekin-17}, the data can be well fit by a piecewise linear function.  Initially, we allowed the slope of both high and low temperature components to be free parameters, as well as the break point where the slope changes.  $\log Q$ is defined in Eqn. \ref{p-md}, and depends on the period and the mean density of the model.  In Figure \ref{piecewise}, we also show the behaviour of these individual components of $Q$ for one of our models. The behaviour of $\log Q$ appears to be dominated by the behaviour of $\log \bar{\rho}$, which shows similar trends to $\log Q$. 

As can be seen in Figure \ref{piecewise}, the shape of the data at high temperature is suggestive of a curved line rather than a straight line as was used in \citet{Lovekin-17}. Although the data can be fit with a linear function in this region, we found the results of our fit in this region were improved by including a second-order term. This effect is especially pronounced at low metallicity, and so we fit the $\log Q-\log T_{eff}$ data with a parabolic function above the break point. 

As found in previous work, the break point is near $\log T_{eff}$ = 3.83. We found that there was no statistically significant variation in the break point with metallicity, rotation rate, or overshoot. However, in some of the lowest metallicity models, the model temperatures are consistently higher than the break point, making it difficult to get a reliable fit in these models. To improve the results, we repeated the fitting procedure while holding the break point fixed at $\log T_{eff} = 3.828$. This improved the fit to the low temperature slope in the low metallicity models, while having no significant effect on the higher metallicity models. However, fixing the break point did affect the coefficients in the fit to the high temperature data. To avoid introducing bias, we chose to keep the break point as a free parameter.

\begin{figure}
    \centering
    \plotone{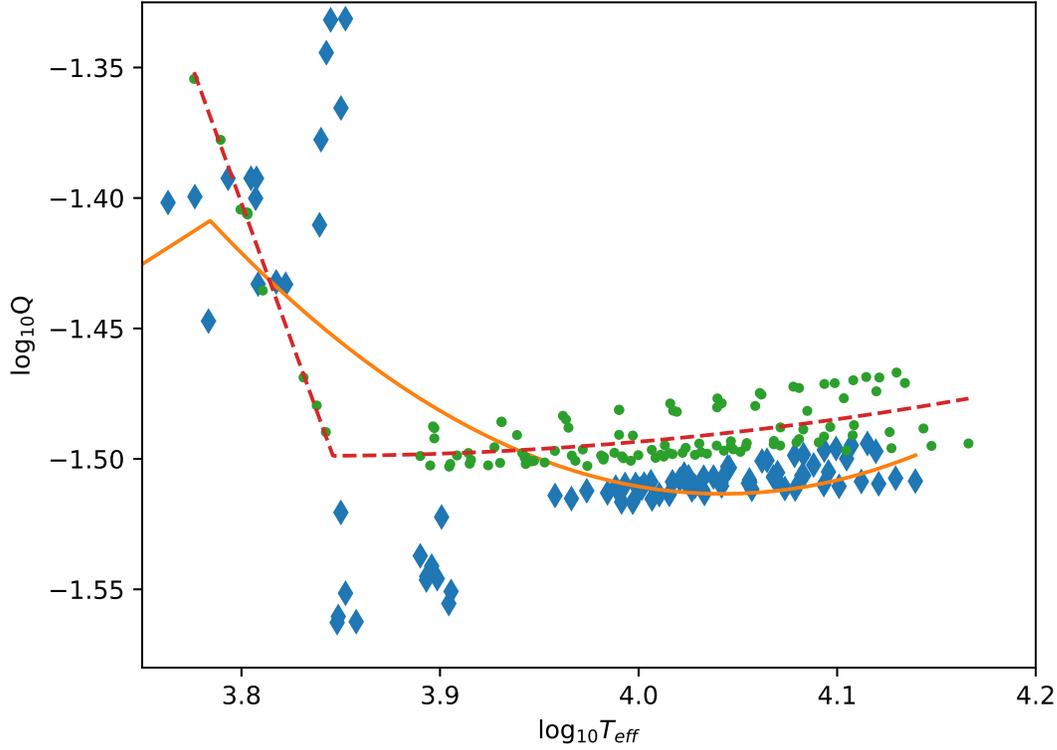}
    \caption{The piecewise fit (solid line) to the $\log Q - \log T$ data (blue diamonds) for the model with $Z = 0.001$, $v/v_{crit} = 0.4$, and $f_{ov} = 0.02$, illustrating how the piecewise fit fails catastrophically. Throughout the main sequence evolution, the model remains at high temperatures, and there is insufficient data to fit a reliable slope for $\log T < 3.83$. For comparison, the model with $Z = 0.001$, $v/v_{crit} =0.1$ is shown for comparison (green circles; dashed line).}
    \label{fig:lowZfit}
\end{figure}

The difficulty in fitting the low metallicity models is most severe in the models with highest rotation, as shown in Figure \ref{fig:lowZfit}. In these models, there are not enough data points to get a reliable fit to the slope at low temperature, whether the break point is held fixed or free. The models with $Z = 0.001$ and $Z = 0.005$ at rotation rates of $v/v_{crit} = 0.3$ and $0.4$ are excluded from the fitting shown here for this reason.

The fitting function used in this paper is 
\begin{equation}
\label{eqn:fitfunc}
    \log Q =\begin{cases}
    m_{low} \log T + T_0 & \log T < T_{break}\\
    C_1 (\log T)^2 + m_{high} \log T + Q_0 & \log T > T_{break}
    \end{cases}
\end{equation}
where the value of $T_0$ is chosen by requiring that the function be continuous at the break point, T$_{break}$. The parameters $m_{high}$ and $m_{low}$ refer to the slope of the fitting function above and below the break point respectively.

\subsection{Low Temperature Results}
\label{sec:m_low}

\begin{figure}
    \centering
    \plotone{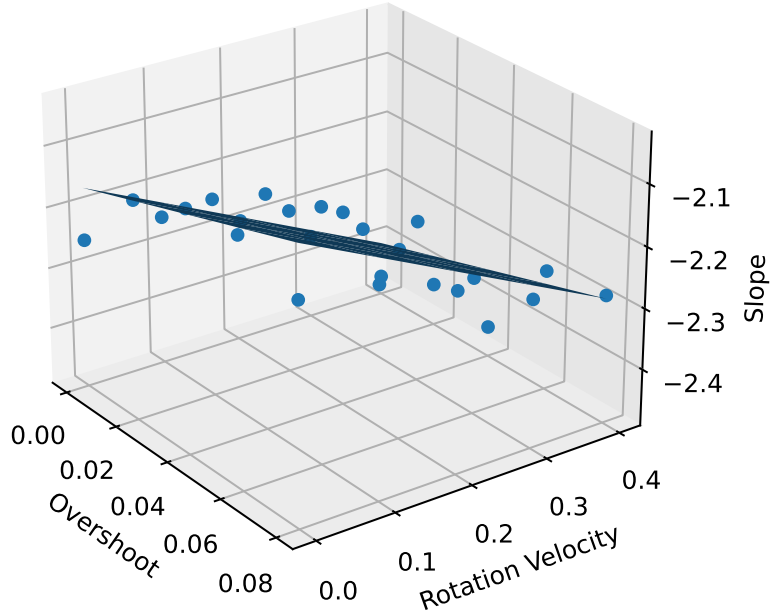}
    \caption{The 3 dimensional fit to the $m_{low}$ data for Z = 0.02 models.  There is some variation with rotation and overshoot, although it is small compared to the scatter in the data.}
    \label{fig:samplefit}
\end{figure}

We then found the behaviour of the low-temperature slope in the piece-wise function, $m_{low}$, As a function of overshoot and rotation rate for each metallicity. To do this, we fit the value of $m_{low}$  using the formula
\begin{equation}
\label{slopefit}
    m_{low} = C_2 f_{ov} + C_3 (v/v_{crit}) + C_4. 
\end{equation}
We found that including metallicity in a 3 parameter fit introduced large errors, and it is therefore more interesting to look at each metallicity individually. As will be discussed below, the behaviour of the fits at each metallicity is quite similar. In this fitting process, the slope values ($m_{low}$) were weighted with the formal uncertainty produced by the fits to the $\log Q - \log T$ data.  The resulting fits are shown in Figure \ref{fig:samplefit} for the Z = 0.02 case. As is typical, there is a large scatter in the model data points.  

\begin{figure}

\plotone{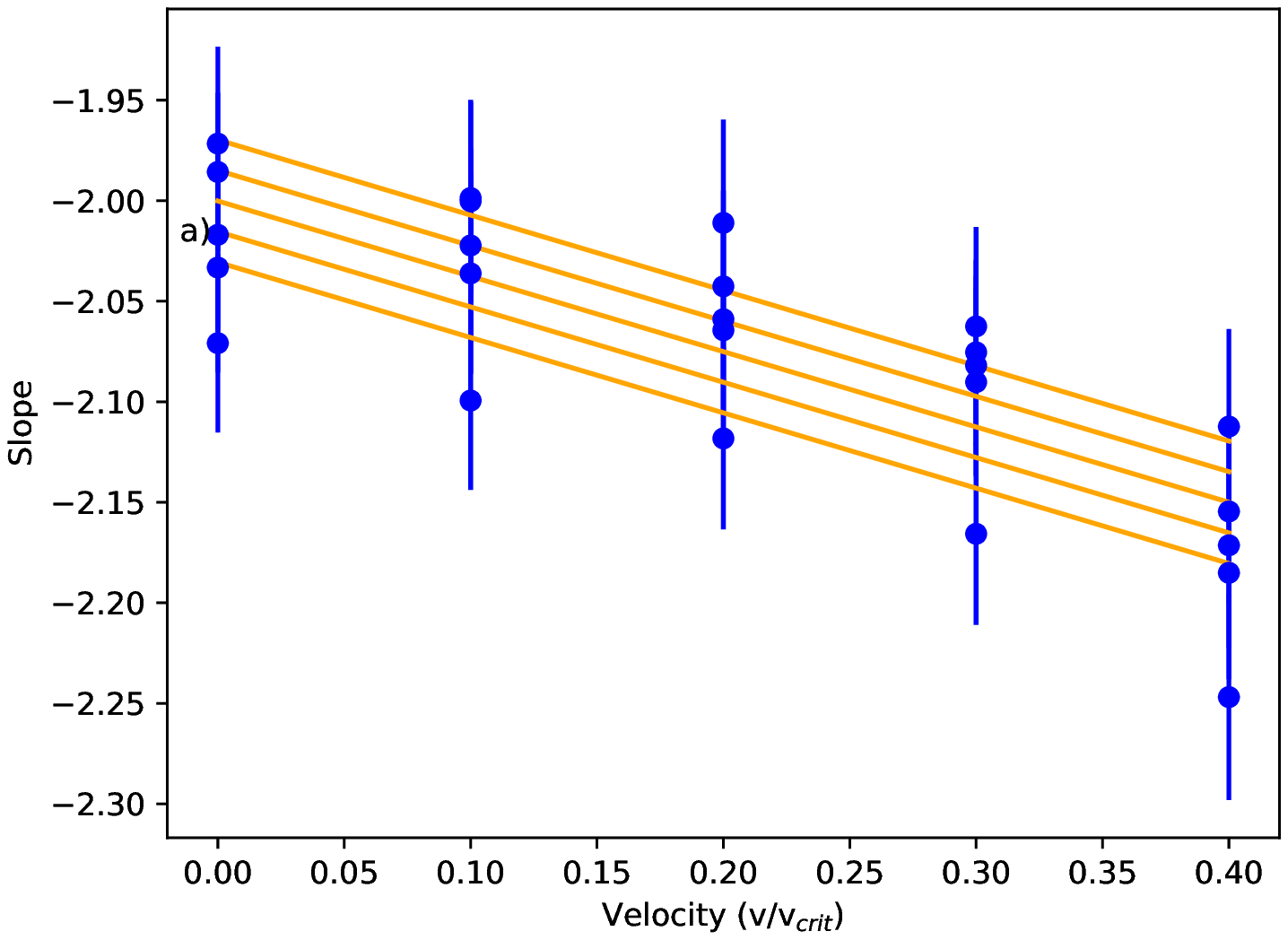}
\plotone{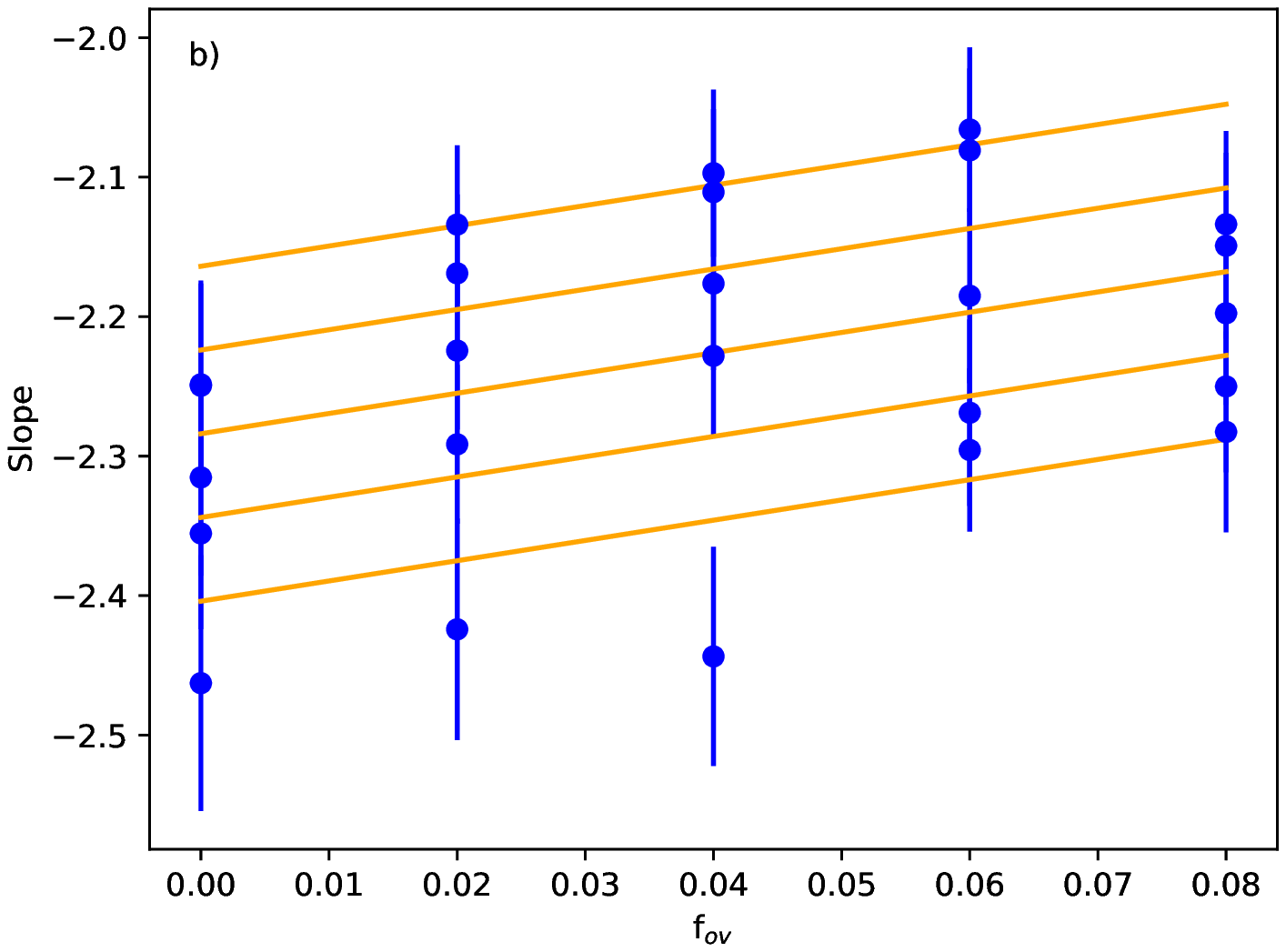}
\caption{\label{fig:slopefits}Fits to the $m_{low}$ data as a function of rotation velocity (a) and overshoot (b).  Data points are shown with formal error bars from fitting the $\log Q - \log T$ data. Lines show result of 3 dimensional fit in Equation \ref{slopefit} projected into 2 dimensions. In part (a), the overshoot increase from the bottom line to the top line. In part (b), the velocity increases from the top line to the bottom line. Data shown is for $Z = 0.02$. }
\end{figure}

To more clearly visualize the effects of the scatter in the models, we show 2D projections of the data for the $Z = 0.02$ models in Figure \ref{fig:slopefits}. In part a), each overshoot is plotted as a function of velocity, and in part b) the slopes at each velocity are plotted as a function of overshoot. Similar results are seen at all metallicities, with the slope $m_{low}$ becoming steeper as the rotation velocity increases and shallower as the overshoot increases. This significantly disagrees with the results of \citet{Lovekin-17}, who found that there was no rotation dependence and a strong dependence on overshoot, given by
\begin{equation}
    m_{low} = (12\pm2)f_{ov} - (2.8\pm 0.1).
\end{equation}
The sense of the overshoot trend is the same here, but the effect is not as strong. 
 
The difference between our results and \citet{Lovekin-17} appears to arise from improvements to the convective boundary treatment in the underlying MESA models. In \citet{Lovekin-17}, the models were made using MESA version 8845, while the models in this work are created using MESA version 12115. In between these two versions, MESA introduced a new ``convective premixing" scheme, as detailed in \citet{Paxton-19}. This new scheme slightly shifts the location of convective boundaries, which in turn affects the size of the oscillation cavity, and hence the pulsation frequency.  

The results presented here seem reasonable based on our understanding of stellar structure. Increased rotation rates in these models have minimal effect on the size and location of the convective core, while increasing the size of the envelope. This effectively increases the size of the oscillation cavity, which reduces $Q$. Since rotating models are generally cooler than a non-rotating model of the same temperature, a rotating model will have a lower $Q$ than a non-rotating model of the same temperature. In terms of fractional radius, the difference decreases as temperature increases, decreasing the slope in $\log Q$ vs. $\log T_{eff}$.

Increased overshoot on the other hand decreases the size of the oscillation cavity by extending both the core and surface convective zones. This increases $Q$, and tending to increase the slope of in $\log Q$ vs. $\log T_{eff}$. 

\begin{deluxetable}{llll}
\label{tab:params}
\caption{Fit Values}
\tablehead{\colhead{Metallicity} & \colhead{$C_2$} & \colhead{$C_3$} & \colhead{$C_4$}}
\startdata
Z = 0.001 & $0 \pm 1 $ & $5.8 \pm 0.9$ & $-2.3 \pm 0.2$\\
Z = 0.005 & $ -9 \pm 2 $ & $-1.8 \pm 0.9 $ & $-2.1 \pm 0.1$ \\
Z=0.01 & $2 \pm 1 $ & $-0.7 \pm 0.4$ & $-2.13 \pm 0.09$\\
Z=0.02 & $1.4 \pm 0.3$ & $-0.60 \pm 0.07$ & $-2.16 \pm 0.02$\\
Z=0.03 & $1.5 \pm 0.4$ & $-0.56 \pm 0.07$ & $-2.05 \pm 0.02$ \\
Z=0.04 & $0.8 \pm 0.2$ & $-0.37 \pm 0.05$ & $-2.03 \pm 0.02$ 
\enddata
\end{deluxetable}

\subsection{High Temperature Results}

\begin{figure}
    \centering
    \plotone{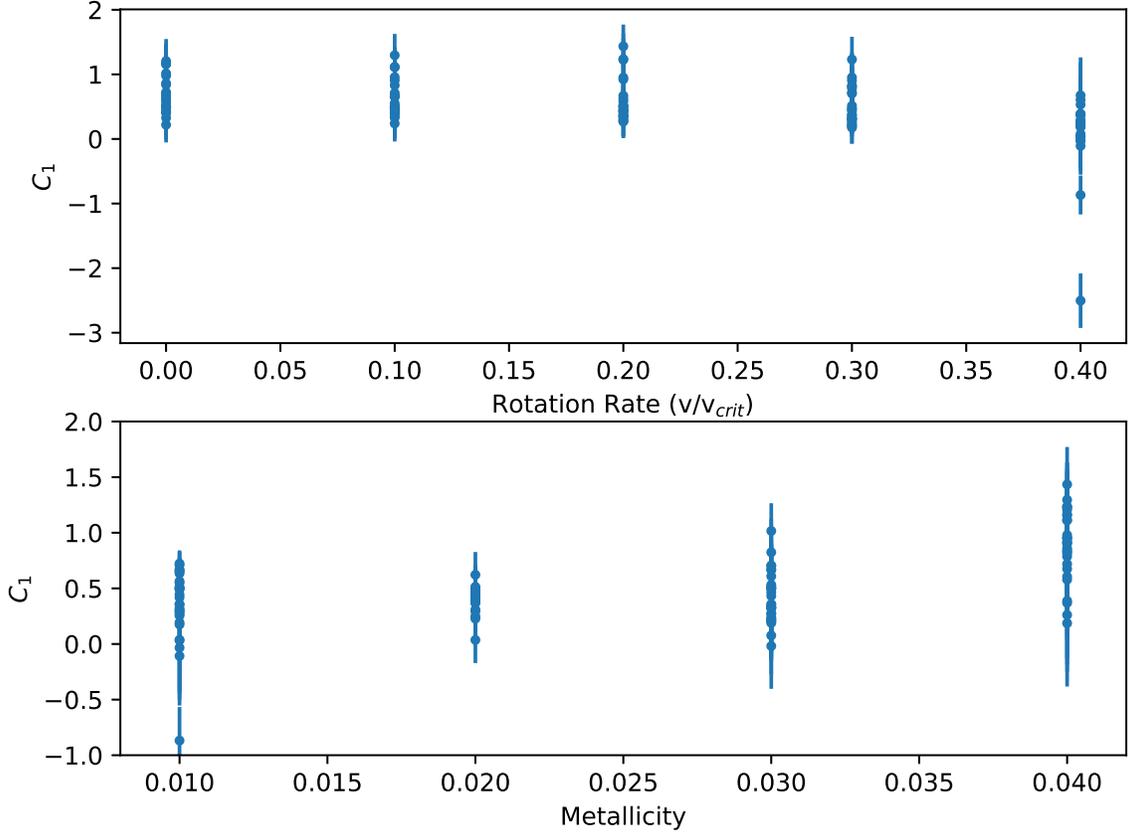}
    \caption{The behaviour of the coefficient $C_1$ as a function of rotation rate (top) and metallicity (botom). The scatter is large enough that there are no significant trends, although there are some indications that $C_1$ decreases slightly at high rotation rates and increases with metallicity. The lowest metallicity models are excluded from this figure.}
    \label{fig:c1trends}
\end{figure}

As discussed previously, the $\log Q$ values are better fit by a parabola at temperatures above the break point. This effect was not previously seen by \citet{Fitch-81} or \citet{Lovekin-17} as the models did  not extend to sufficiently high temperatures.  We examined the behaviour of the coefficient $C_1$ in Equation \ref{eqn:fitfunc} as a function of convective core overshoot, rotation rate, and metallicity.  The results are shown for rotation and metallicity in Figure \ref{fig:c1trends}.

We found that there was no significant trend in the value of $C_1$ as a function of metallicity, rotation rate, or convective core overshoot. There are indications of possible trends with rotation and metallicity, but the scatter is too large to determine a significant result. Instead, the parabolic shape can be explained by the changing size of the oscillation cavity, as discussed in Section \ref{sec:convection}.

\subsection{Metallicity Effects}

\begin{figure}
    \centering
    \plotone{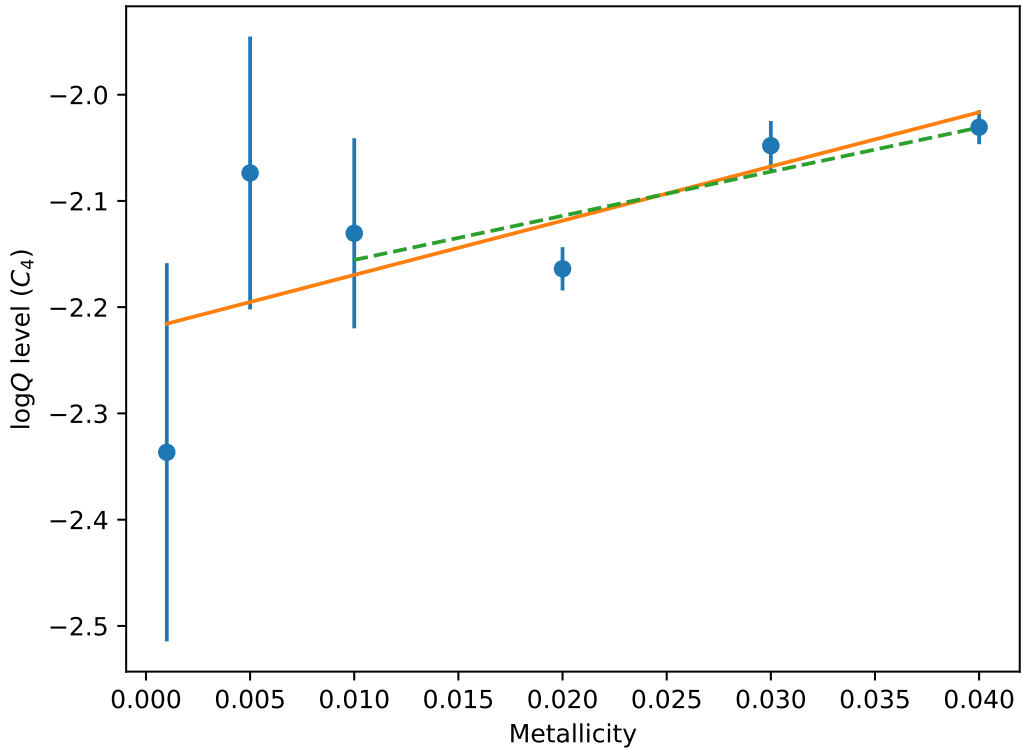}
    \caption{The metallicity dependence of the parameter $C_4$ as a function of metallicity. The solid line shows the fit to the full data set (Eqn. \ref{eqn:allZ}), while the dashed line shows the fit when the low metallicity data excluded (Eqn. \ref{eqn:metaleffect}). Error bars show the formal errors on the fitting taken from Table \ref{tab:params}.}
    \label{fig:zslopes}
\end{figure}

In addition to the dependence on rotation rate or convective overshoot, we investigate a metallicity dependence, focusing on the parameter $C_4$ (column 4 of Table 2).
Overall, there is a slight increase in $C_4$ as a function of metallicity. We are able to fit the $C_4$ data with a simple linear regression, which yields a best fitting line of 
\begin{equation}
\label{eqn:allZ}
    C_4 = (5 \pm 3) Z - (2.22 \pm 0.06).
\end{equation}
It must be noted that the uncertainty on $C_4$ is very large for the $Z = 0.001$ and Z = 0.005 models.  If we exclude these model as outliers, the error on the slope becomes  smaller, and the best fit line is given by 
\begin{equation}
\label{eqn:metaleffect}
    C_4 = (4 \pm 2) Z - (2.20 \pm 0.05).
\end{equation}
Both results are quite similar, and show a slight upward trend with metallicity, as shown in Figure \ref{fig:zslopes}. Although the metallicity trend shown here is significant given the formal errors on the fits in Table \ref{tab:params}, the large scatter in the data means this finding is unlikely to be usable in practice.

\begin{figure}
    \centering
    \plotone{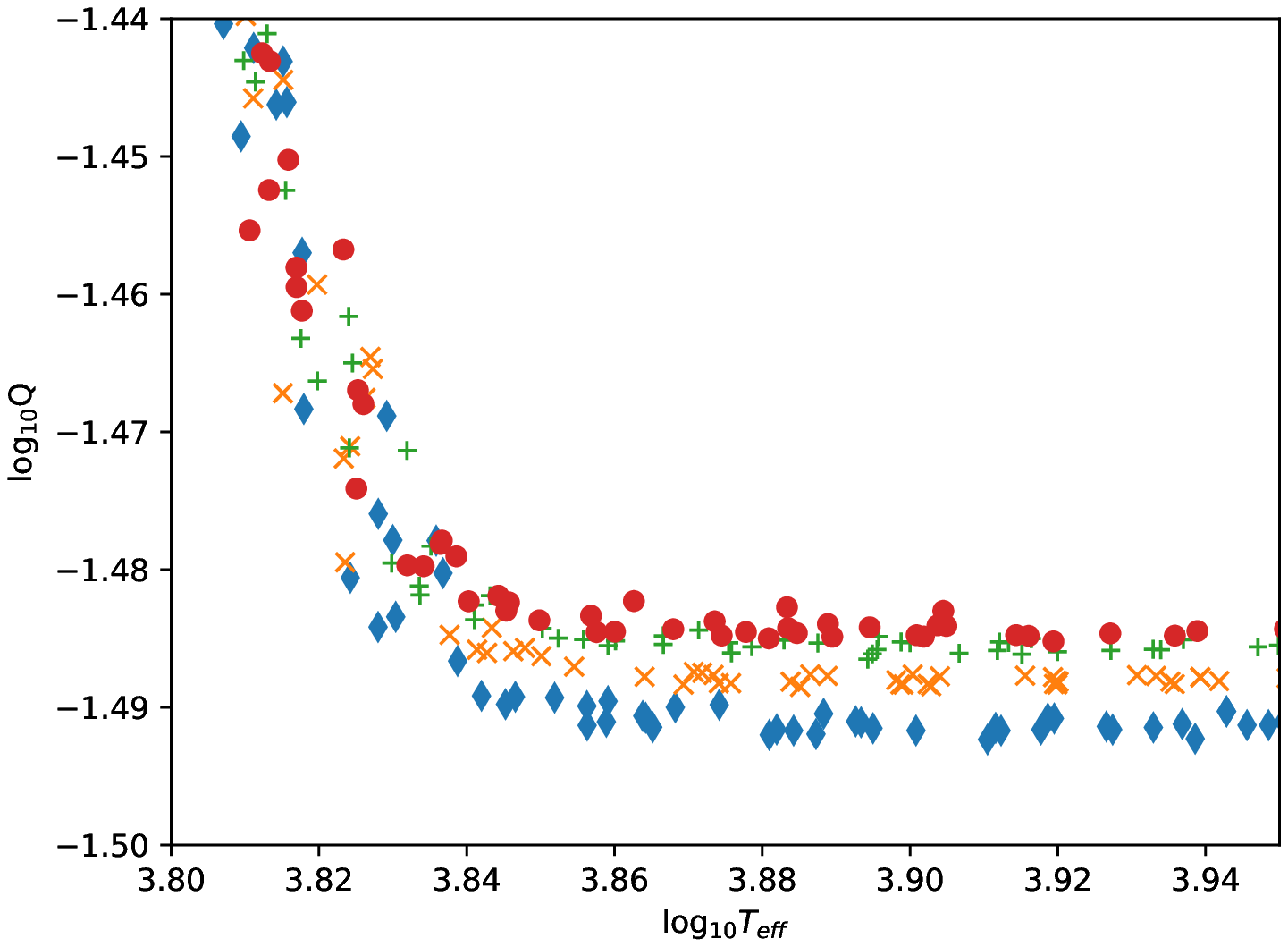}
    \caption{The $\log Q$ vs $\log T_{eff}$ behaviour for non-rotating models with no overshoot for metallicities Z = 0.01 ($\Diamond$), 0.02 (x), 0.03 (+), and 0.04 ($\circ$). The trend to larger $\log Q$ values with metallicity can be clearly seen.}
    \label{fig:Zeffect}
\end{figure}

The increase in $C_4$ corresponds to a global increase in the value of $\log Q$ at the break point (parameter $Q_o$ in Equation \ref{eqn:fitfunc}) as illustrated in Figure \ref{fig:Zeffect} for Z = 0.01, 0.02, 0.03, and 0.04 for models with no overshoot or rotation.  
The increase in $\log Q$ as a function of metallicity is to be expected, as lower metallicity stars are generally hotter, more luminous, and more centrally condensed than their higher-metallicity counterparts \citep{Hirschi2008}.  This is expected to produce shorter periods and higher average densities. Based on our results, the increase in average density offsets the shorter periods, producing an increase in $\log Q$ as seen in Figure \ref{fig:Zeffect}. 

\section{Influence of Convection}
\label{sec:convection}

\cite{Lovekin-17} found that the location of the break point in the $\log Q- \log T_{eff}$ plot was determined by the location of the surface convection zones, which begin to penetrate further into the star for models cooler than $\log T_{eff}$ = 3.83 (see their Figure 3). In this work, we have confirmed that this is true for models at all metallicities.  

\begin{figure}
    \centering
    \plotone{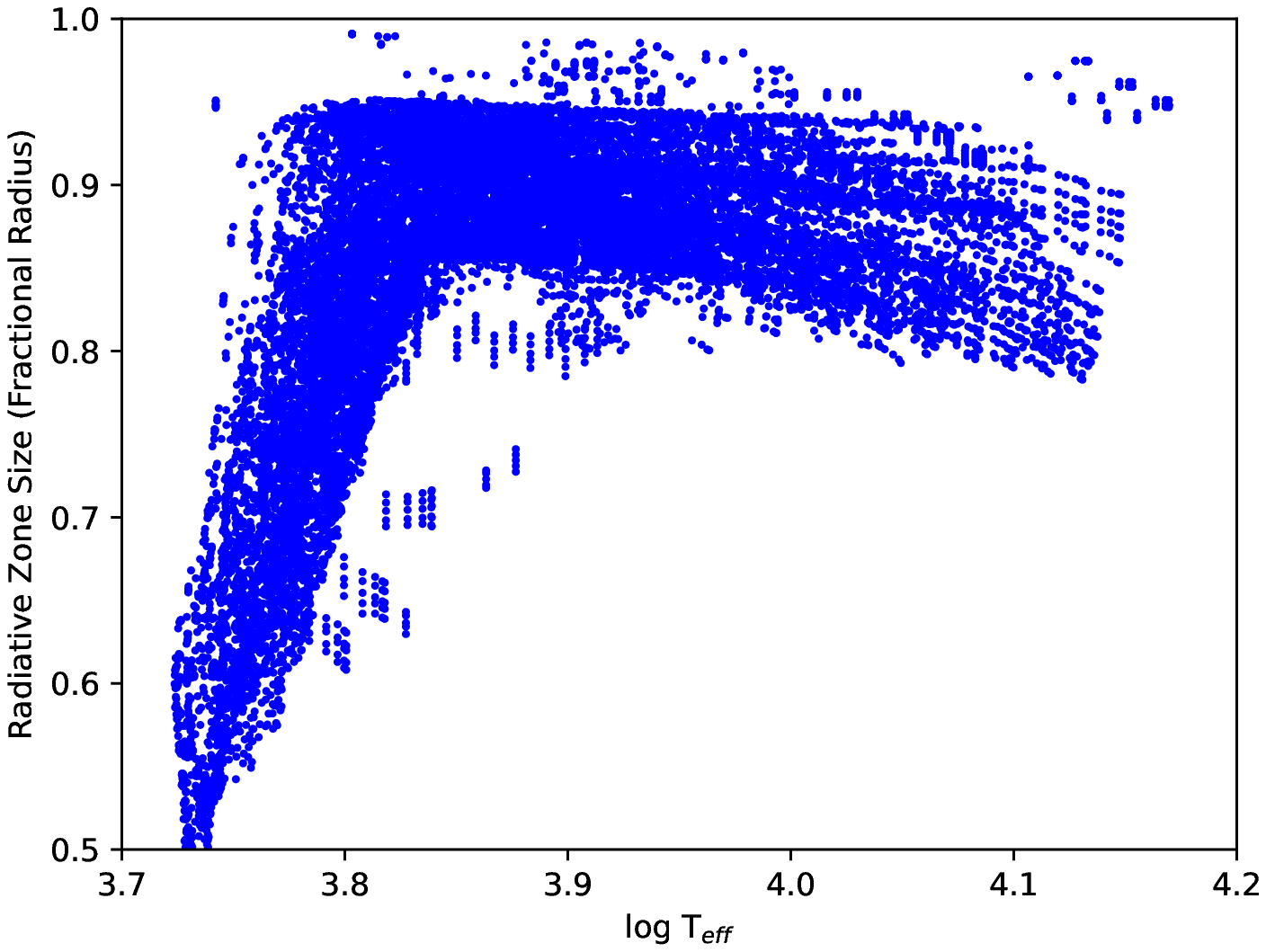}
    \caption{The fractional size of the oscillation cavity in each model as a function of effective temperature. The size of the cavity is determined by the boundary of the convective core and the location of the innermost surface convection zone. The change in slope below $\log T_{eff} = 3.83$ is clearly visible, as is the parabolic behaviour seen in Figure \ref{piecewise}.}
    \label{fig:cavity}
\end{figure}

The size of the oscillation cavity is determined by the location of both the surface convection zone (if present) and the size of the convective core. In Figure \ref{fig:cavity}, we plot the size of the oscillation cavity, determined as the distance between the innermost and outermost convective zones, as a fraction of the total radius of the star as a function of temperature. Again, the sharp change in slope at temperatures below $\log T_{eff} = 3.83$ is clearly visible. In addition, it can be seen that the size of the oscillation cavity starts to decrease in the hottest models in our sample, mirroring the parabolic behaviour of $\log Q-\log T_{eff}$ seen in Figure \ref{piecewise}. Indeed, this decrease in the size of the oscillation cavity corresponds to a relative increase in convective core size starting around $\log T_{eff} = 3.9$. 

Based on these results, we can confirm that the behaviour of $\log Q$ as a function of $\log T_{eff}$ is dependent on the size of the radiative cavity between the convective core and the surface convection zones. The change in slope at the break-point corresponds to the temperature at which the deepest surface convection zone moves closer to the surface of the star. As was found by \cite{Lovekin-17}, the depth of the surface convective zone changes the sound speed gradient near the surface, which will affect the frequency, and hence the pulsation constant. 

\section{Conclusion} 
\label{sec:conclusion}

We have expanded on the work of \citet{Lovekin-17} by updating and extending the models to cover a range of metallicities. Models of $\delta$ Scuti variable stars with varying masses, rotation rates, convective core overshoot parameter, and chemical compositions was created to investigate the behaviour of the pulsation constants of the models as a function of their effective temperatures in relation to the amount of convective overshooting. These updated models find a similar correlation between $m_{low}$ and convective core overshoot, although the relationship is not as strong. In addition, we find a correlation between $m_{low}$ and rotation rate, which was not found in previous work. The results were consistent across all metallicities. 

A slight trend with metallicity was seen in the constant $C_4$ in Equation \ref{slopefit}, and the entire curve shifts to higher $\log Q$ values as the metallicity increases. Combining all these results, the slope of $\log Q$ vs $\log T_{eff}$ can be predicted by Equation \ref{eqn:metaleffect} when  $\log T_{eff} < 3.83$.

By extending our model grid to higher temperatures, we find an upward trend in $\log Q$ at high $T_{eff}$, and the $\log Q$ values are better fit by a parabola above $\log T_{eff} \sim 3.83$. This upward trend is produced by increasing convective core size, reducing the size of the oscillation cavity.  The value of this fit parameter ($C_1$) was not found to be strongly dependent on rotation rate, convective overshoot parameter, or metallicity.

We have also confirmed that the behaviour of $\log Q$ is connected to the location and size of the central and surface convection zones in the star.  As shown in Figure \ref{fig:cavity}, the size of the radiative cavity closely mirrors the behaviour of $\log Q$ when both are shown as a function of $\log T_{eff}$. In the hottest models in our sample, the convective core is larger than in the cooler models, shrinking the radiative cavity. The smaller radiative cavity increases Q, producing a parabolic upturn in $\log Q$ as a function of $\log T_{eff}$.

\acknowledgments The authors gratefully acknowledge support from the Natural Sciences and Engineering Research Council of Canada.  This research has made use of NASA's Astrophysics Data System.
\software{MESA (Paxton et al., 2011), GYRE (Townsend \& Teitler 2013), NumPy (Harris et al., 2020), SciPy (Virtanen et al., 2020),  Matplotlib (Hunter 2007)}

\vspace{5mm}
\bibliography{Main}

\end{document}